\documentclass[11pt,titlepage]{article}
\usepackage{setspace}
\usepackage{amsmath}
\usepackage{graphicx}
\usepackage[round,numbers,sort&compress]{natbib}
\begin{document}

\title{Neuronal synchrony during an{\ae}sthesia - A thalamocortical model}

\author{Jane H.~Sheeba\\
Department of Physics,\\ Lancaster University,\\ Lancaster, LA1
4YB, UK \and Aneta~Stefanovska \thanks{Department of Physics,
Lancaster University, Lancaster, LA1 4YB, UK}\\
Department of Physics,\\ Lancaster University,\\ Lancaster, LA1
4YB, UK \and Peter~V.~E.~McClintock\\ Department of Physics,\\
Lancaster University,\\ Lancaster, LA1 4YB, UK}

\date{\today}

\begin{abstract}

There is growing evidence in favour of the temporal-coding
hypothesis that temporal correlation of neuronal discharges may
serve to bind distributed neuronal activity into unique
representations and, in particular, that $\theta$ (3.5-7.5 Hz)
and $\delta$ ($0.5<$3.5 Hz) oscillations facilitate information
coding.
The $\theta$ and $\delta$ rhythms are shown to be
involved in various sleep stages, and during an{\ae}sthesia,
and they undergo changes with the depth of an{\ae}sthesia.
We introduce a thalamocortical model of interacting neuronal
ensembles to describe phase relationships between $\theta$ and
$\delta$ oscillations, especially during deep and light
an{\ae}sthesia. Asymmetric and long range interactions among
the thalamocortical neuronal oscillators are taken into
account. The model results are compared with the experimental
observations of Musizza et al. {\it J.\ Physiol. (London)} 2007
580:315-326. The $\delta$ and $\theta$ activities are found to
be separately generated and are governed by the thalamus and
cortex respectively. Changes in the degree of intra--ensemble
and inter--ensemble synchrony imply that the neuronal ensembles
inhibit information coding during deep an{\ae}sthesia and
facilitate it during light an{\ae}sthesia.

\end{abstract}

%

\maketitle
\section{Introduction}

Neuronal communication and synchronization are crucially
important features of the co--operative interaction between
neuronal ensembles, endowing the brain with the marvelous
capability known as \emph {cognition}. The time
scales of human motor and cognitive events are comparable to
that of the EEG field dynamics arising from synchronous
activity, observable from within or outside the brain
\cite{Negro:02}. Since EEG signals correspond to the averaged
activity of large cell populations, their fluctuations can
support cognitive activity only if they are sufficiently
synchronized: effective communication of neuronal ensembles can
be achieved only if they are oscillating in a synchronized
manner. Such behavior tags those neurons to represent
particular cognitive tasks, e.g.\ in relation to a perceptual
object \cite{Singer:99}.

In this article we present a model of interacting
thalamocortical neuronal ensembles in an attempt to account for
the behaviour of $\delta$ and $\theta$ waves during
an{\ae}sthesia. Our model is motivated by an attempt to tackle
the problem of an{\ae}sthetic awareness, to provide the basic
understanding needed to prevent people inadvertently awakening
during surgery. We therefore focus on ways to identify and
characterize the states of deep and light an{\ae}sthesia. Our
starting point model is the experimental observation that both
the cardio-respiratory and respiratory-$\delta$ interactions
change with depth of an{\ae}sthesia in rats
\cite{Musizza:07,Stefanovska:00a} following the administration
of a single bolus of ketamine-xylazine. During the ensuing deep
phase of an{\ae}sthesia ($\sim$45 min) the amplitude of
$\delta$-waves is strongly pronounced. As the subsequent light
phase of an{\ae}sthesia ($\sim$25 min) is entered, the
$\delta$-waves disappear and the amplitude of $\theta$-waves
increases. We use the phase dynamics approach to propose a
model that incorporates regional ensembles of neurons that are
connected both among and between themselves.

Further, we investigate the role played by the neuronal
ensembles in temporal coding during deep and light
an{\ae}sthesia from the model results. Neuronal synchrony is
often associated with an
oscillatory pattern of signals (oscillation--based synchrony).
The frequencies of such signals generally cover a broad range
and, more importantly, exhibit a marked state dependence. In
other words, synchrony and temporal coding of information in
selective frequency bands are two different sides of the same
coin. We use this idea and establish the link between the conditions
under which temporal coding is believed to occur (synchrony) and
the level of arousal of the brain.

\subsection{Physiological background}
During periods of slow wave sleep/an{\ae}sthesia, widespread
synchronized oscillations occur throughout the thalamus and the
cortex. The slow oscillations ($<1$Hz) that occur during
natural sleep and ketamine--xylazine an{\ae}sthesia are an
emergent network property of neocortical neurons involving the
TC and RE neurons \cite{Contreras:97}. The clock--like $\delta$
($1-4$Hz) are produced in the thalamus and are strongly
synchronized by the RE and the thalamocortical volleys
\cite{Bal:96}.
A small proportion of the TC neurons also display slow and
$\delta$ oscillations \cite{Contreras:97}. Thus the high
amplitude, low frequency $\delta$ and slow oscillations are
found to be associated with highly coherent activities of the
cortical, RE and TC neurons. The dynamics of $\delta$ and
$\theta$ waves can therefore be represented by interacting
ensembles of neuronal oscillators.

\begin{figure}
\begin{center}
\includegraphics[angle=0, width=5.5cm]{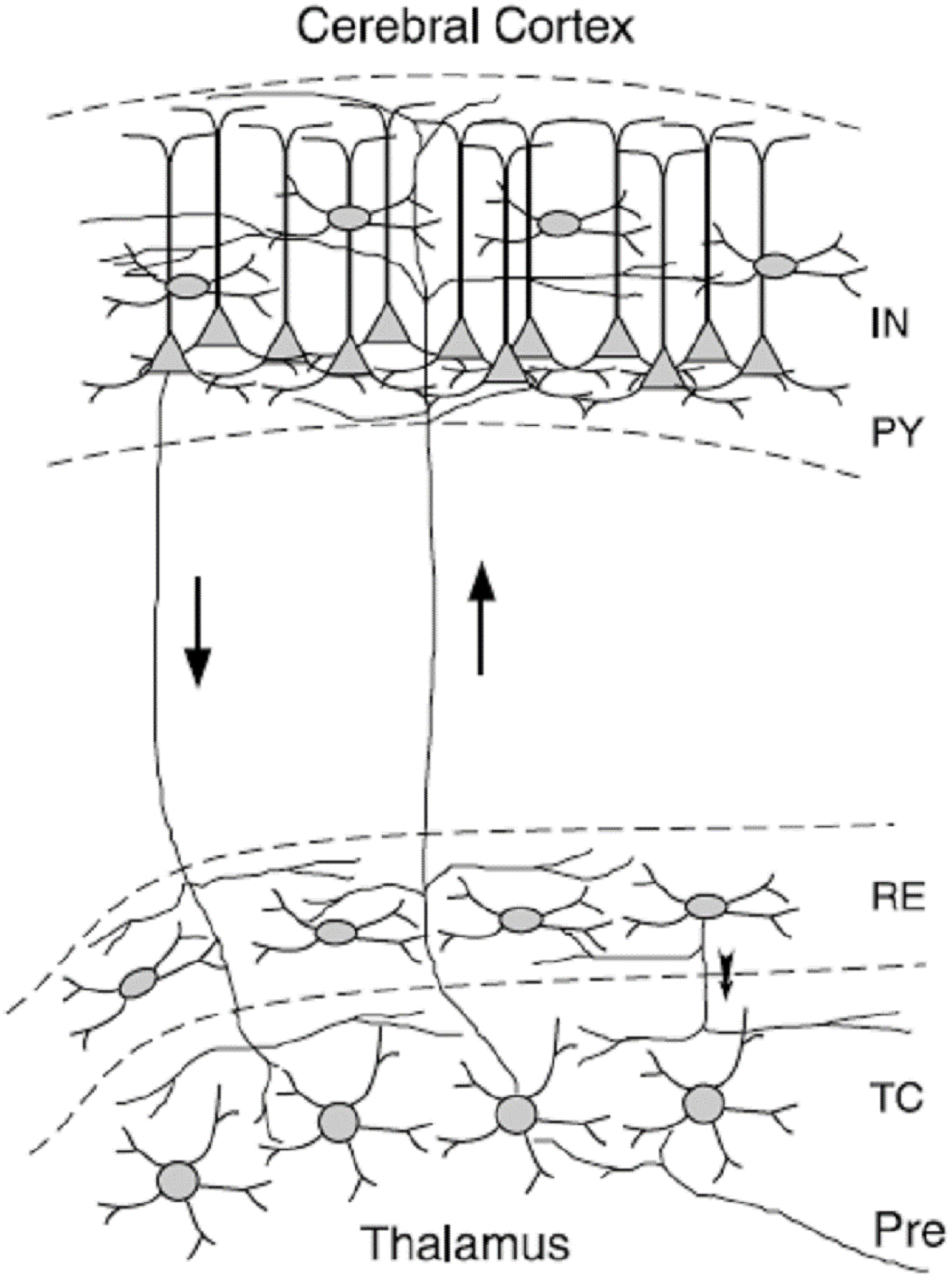}
\includegraphics[angle=0, width=6.5cm, height=7cm]{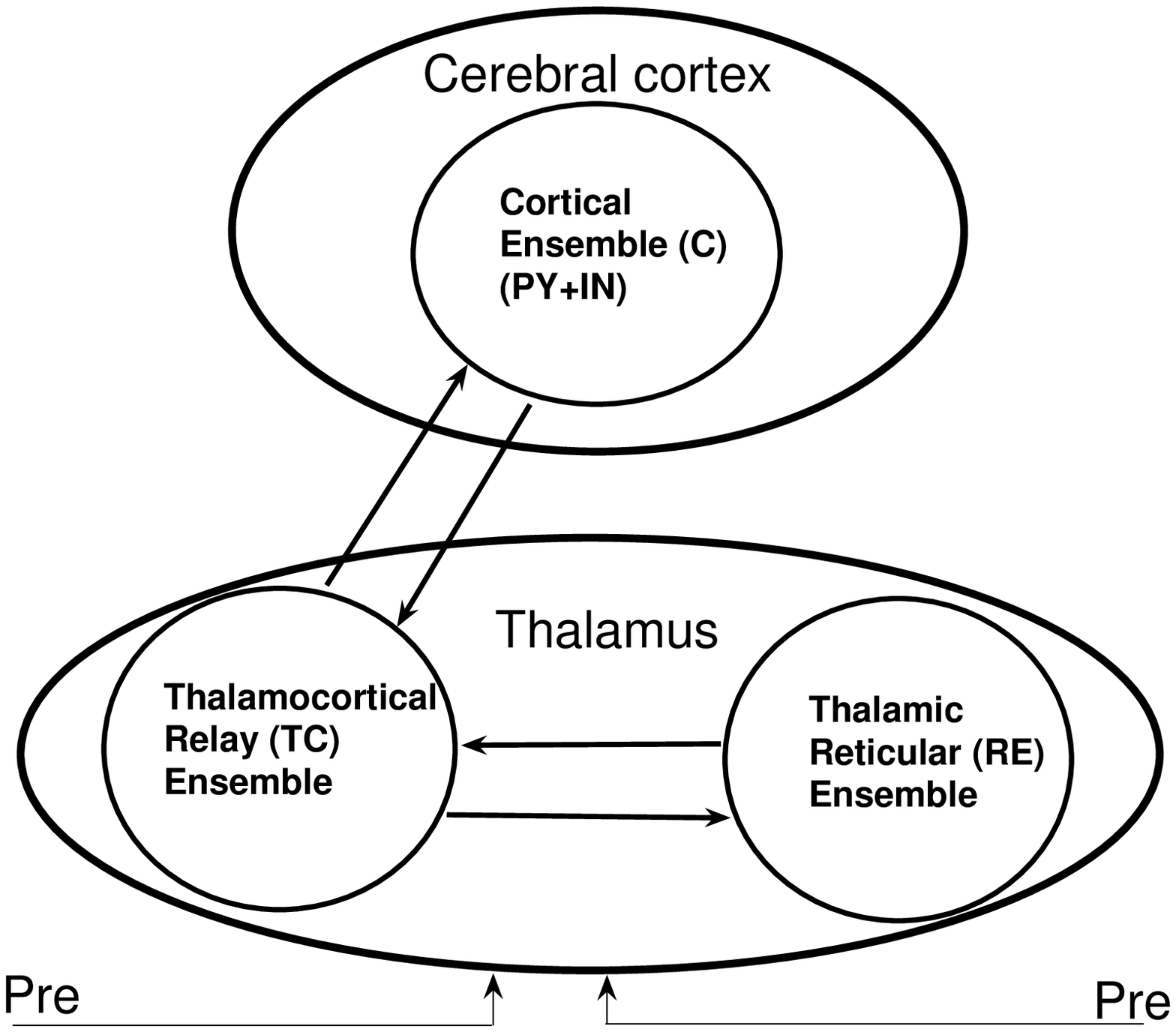}
 \caption{}
\label{schematic}
\end{center}
\end{figure}

The physiological circuitry is as shown in Fig. \ref{schematic}
(left) \cite{Destexhe:03}. The pyramidal (PY) neurons of the
cerebral cortex are connected among themselves (excitatory) by
the interneurons (IN) which also form components of the
cortical ensemble. The TC neurons receive sensory inputs which
are relayed to the appropriate area of the cortex through
ascending thalamocortical fibres (indicated by the upward
arrow). The RE neurons wrap most of the dorsal and ventral
aspects of thalamus \cite{Jones:85} and act as bridge between
the TC neurons and the thalamic neurons. The dense axons of the
RE neurons innervate the TC neurons \cite{Cox:96}. The
corticothalamic fibers (indicated by the downward arrow) also
leave collaterals within the RE nucleus and dorsal thalamus.
The RE neurons thus form a network that surrounds the thalamus.
It receives a copy of nearly all thalamocortical and
corticothalamic activity, and projects connections \emph
{solely} to neurons in the TC region \cite{Destexhe:03}. In
turn, the axons of the TC neurons give rise to collaterals in
the RE nucleus while the parent axon passes through the
cerebral cortex \cite{Harris:87}.

\section{The model}

Based on the physiological phenomena that take place during
an{\ae}sthesia, as well as on anatomy, the model system
considered is shown schematically in Fig.\ \ref{schematic}
(right). The oscillators in the three ensembles, namely the
cortical, the thalamic reticular and the thalamocortical relay
neurons, have different mean natural frequencies and their
interactions are characterized by intra-population and
inter-population coupling parameters.

Each neuron in the ensembles is considered as an oscillator
whose membrane potential is the oscillating variable. The
couplings represent the synaptic connections between them. We
reduce the system to a phase model, one of the simplest yet
accurate models for weakly coupled nonlinear oscillators, with
the coupling being introduced through the phases. In doing so,
we make use of the fact that there exists a degree of coherence
between the membrane potential oscillations and the action
potential firings. Although they are not equivalent, the action
potentials are triggered at a certain phase of the membrane
potential oscillations. This reasoning leads to the following
set of equations
\begin{eqnarray}
\label{rmod01}
 \dot{\theta_i^{(1)}}&=& \omega_i^{(1)} -
 \frac{A_c}{N}\sum_{j=1}^{N}
\sin(\theta_i^{(1)}-\theta_j^{(1)}+\alpha) \nonumber
\\&&\quad\quad -
\frac{B_c}{N}\sum_{j=1}^{N}\sin(\theta_i^{(1)}-\theta_j^{(2)}+\alpha)+\eta_i^{(1)},
 \nonumber\\
\dot{\theta_i^{(2)}}&=&\omega_i^{(2)}-\frac{A_{tc}}{N}\sum_{j=1}^{N}
\sin(\theta_i^{(2)}-\theta_j^{(2)}+\alpha) \nonumber
\\ &&\quad\quad - \frac{B_{tc}}{N}\sum_{j=1}^{N}
\sin(\theta_i^{(2)}-\theta_j^{(1)}+\alpha)
\\ &&\quad\quad - \frac{C_{tc}}{N}\sum_{j=1}^{N}
\sin(\theta_i^{(2)}-\theta_j^{(3)}+\alpha)+\eta_i^{(2)}, \nonumber \\
\dot{\theta_i^{(3)}}&=& \omega_i^{(3)} -
\frac{A_{re}}{N}\sum_{j=1}^{N}
\sin(\theta_i^{(3)}-\theta_j^{(3)}+\alpha) \nonumber \\
&&\quad\quad -
\frac{B_{re}}{N}\sum_{j=1}^{N}\sin(\theta_i^{(3)}-\theta_j^{(2)}+\alpha)+\eta_i^{(3)},
\nonumber
\end{eqnarray}
where $\theta_i^{(1,2,3)}$ are the phases of the $i$th
oscillator in the C, TC and RE ensembles, respectively, and $N$
refers to the ensemble sizes. The parameters $A_c$, $A_{tc}$
and $A_{re}$ quantify intra-ensemble couplings and $B_c$,
$B_{tc}$, $C_{tc}$ and $B_{re}$ quantify the inter-ensemble
couplings. The natural oscillator frequencies
$\omega_i^{(1,2,3)}$ are assumed to be distributed with central
frequencies $\bar\omega^{1,2,3}$. The $\eta_i^{(1,2,3)}$ are
independent white Gaussian noises for which $\langle
\eta_i^{(1,2,3)}(t) \rangle =0$ and $\langle
\eta_i^{(1,2,3)}(t)\eta_j^{(1,2,3)'}(t)\rangle =
2D_{(1,2,3)}\delta(t-t')\delta_{ij}$; $D_{(1,2,3)}$ are the
noise intensities. For convenience one can define
complex-valued, mean field, order parameters as
$r_{(1,2,3)}e^{i\psi_{(1,2,3)}}=\frac{1}{N}\sum_{j=1}^N
e^{i\theta_j^{(1,2,3)}}$. Here $\psi_{1,2,3}(t)$ are the
average phases of the oscillators in the respective ensembles
and $r_{1,2,3}(t)$ are measures of the coherence of the
oscillator ensembles, which vary from 0 to 1. With these
definitions, Eqs.\ (\ref{rmod01}) become
\begin{eqnarray}
\label{rmod02} \dot{\theta_i^{(1)}}&=& \omega_i^{(1)} -
 r_1A_c
\sin(\theta_i^{(1)}-\psi_1+\alpha) \nonumber
\\&&\quad\quad -
r_2B_c\sin(\theta_i^{(1)}-\psi_2+\alpha)+\eta_i^{(1)},
 \nonumber\\
\dot{\theta_i^{(2)}}&=&\omega_i^{(2)}-r_2A_{tc}\sin(\theta_i^{(2)}-\psi_2+\alpha)
\nonumber
\\ &&\quad\quad - r_1B_{tc}\sin(\theta_i^{(2)}-\psi_1+\alpha)
\\ &&\quad\quad - r_3C_{tc}\sin(\theta_i^{(2)}-\psi_3+\alpha)+\eta_i^{(2)}, \nonumber \\
\dot{\theta_i^{(3)}}&=& \omega_i^{(3)} - r_3A_{re}\sin(\theta_i^{(3)}-\psi_3+\alpha) \nonumber \\
&&\quad\quad -
r_2B_{re}\sin(\theta_i^{(3)}-\psi_2+\alpha)+\eta_i^{(3)}.
\nonumber
\end{eqnarray}
\noindent

\subsection{Numerical methods}

A fourth-order Runge-Kutta routine is used for the numerical
simulation with the initial phases equally distributed within
$[0,2\pi]$. The results are normalized to ``real time" and the
system is simulated for the equivalent of 60 mins with
$N=10000$. We will refer to synchrony within an ensemble as
{\it intra--ensemble synchrony}, and that between ensembles as
{\it inter--ensemble synchrony}. The amount of intra--ensemble
synchrony is measured by the mean field parameters:
$r_{(1,2,3)}=0$ implies that there is no synchrony in the
corresponding ensemble; $r_{(1,2,3)}=1$ indicates complete
synchrony; and $0<r_{(1,2,3)}<1$ corresponds to partial
synchrony. The greater the value of $r_{(1,2,3)}$, the more
oscillators are oscillating in synchrony
\cite{Kuramoto:84,Pikovsky:01}. On the other hand,
inter-ensemble synchrony occurs when oscillators from two
different ensembles entrain to a frequency/phase window. In
general, inter-ensemble synchrony can be quantified by a
constant difference in the mean phases $\psi_{(1,2,3)}$.
However, this measurement will be valid only if both the
ensembles are completely locked to each other. For partial
synchrony, the difference in the mean phase will be
oscillating, showing no synchrony. Hence we identify
inter-ensemble synchrony by looking at the time evolution of
the ensemble-averaged frequencies. Also, for increasing
inter-ensemble coupling parameters, a decrease in $r_{(1,2,3)}$
is a signature of inter-ensemble synchrony \cite{Jane:08}.

The inter--ensemble and intra--ensemble coupling parameters are
swept linearly in time in order to mimic the effect of
decreasing concentration of an{\ae}sthetic agent. This
corresponds to the ability of an{\ae}sthetics to affect
thalamocortical signalling, as is well recognized from {\it in
vivo} electrophysiological work on animals \cite{Alkire:00}.
Thus, during deep an{\ae}sthesia, the neuronal oscillators are
very weakly coupled so that they do not interact much (due to
the an{\ae}sthetics blocking the signalling pathways). As the
an{\ae}sthetic concentration wears off the signalling pathways
become unblocked and the couplings between the neuronal
oscillators and ensembles become stronger. Hence we can
simulate the transition from the deeply to the lightly
an{\ae}sthetized state by starting with a very weak coupling
strength (characterizing deep an{\ae}sthesia) and continuously
sweeping them to stronger values with time. The starting values
of the coupling parameters are given in Fig.\ \ref{compare_F1}.
We introduce asymmetry in the model by a phase shift
$0\leq\alpha<\pi/2$. For the values of the mean frequencies we
follow Amzica et al. \cite{Amzica:98,Amzica:02}; $\bar
\omega^{(1)} = 3$~Hz, $\bar \omega^{(2)} = 1.5$~Hz and $\bar
\omega^{(3)} = 1$~Hz, with Lorentzian distributions
$g(\omega^{(1,2,3)}) =
\frac{\gamma}{\pi}(\gamma^2+(\omega-\bar{\omega}^{(1,2,3)})^2)^{-1}$.

Our choice of values for the mean frequencies reflects the
intrinsic properties of the neurons. The cortical neurons are
endowed with intrinsic properties which could be reflected in
field potential recordings as $\delta$ activities. The
frequencies of these faster oscillations evolve around the
upper limit of the delta band (mainly 3--4 Hz)
\cite{Steriade:06,Amzica:98,Amzica:02}. The clock--like
$\delta$ oscillations (1--4 Hz) are mainly produced in the RE
neurons \cite{Contreras:97}. Individual TC neurons are capable
of producing Ca$^{2+}$ spikes and associated action potentials
at frequencies of 0.5-4 Hz \cite{Bal:96}.

\section{Results}

The time evolution of the mean frequency of each ensemble is
plotted in Fig.\ \ref{compare_F1} right. For comparison, the
experimental $\delta$ and $\theta$ frequencies, obtained by
wavelet analysis of the EEG signals from an{\ae}sthetised rats
\cite{Musizza:07}, are plotted in Fig.\ \ref{compare_F1}
(left). In the latter experiments, the $\delta$ activity was
found initially to be of higher amplitude compared to that of
the $\theta$ waves. In particular, during the deep phase of
an{\ae}sthesia, there occurs a strong, high amplitude, $\delta$
activity which greatly diminishes on entry to the light phase
(at $\sim$ $45$ min). The $\theta$ activity runs independently
throughout the whole period of an{\ae}sthesia, but at a much
lower amplitude compared to that of the $\delta$ waves. Note
the differences in the amplitudes corresponding to the color
codes in Fig. \ref{compare_F1} left, which makes the $\theta$
activity visible in the top panel.

The model results suggest that it is the TC and RE ensembles
that generate the $\delta$ activity during deep an{\ae}sthesia.
This $\delta$ is of high amplitude due to the strong synchrony
in and between the TC and RE ensembles. Synchrony between the
TC and RE ensembles can be easily established even with a
smaller value of inter--ensemble coupling, because the
frequency difference between them is relatively small. Fig.\
\ref{compare_R1} shows the time evolutions of the mean field
parameters $r_{(1,2,3)}$ of the three ensembles corresponding
to the simulated frequencies. During deep an{\ae}sthesia, the
fractions of oscillators that oscillate in synchrony in the TC
and RE ensembles are higher than in the C ensemble. This
results in a strongly pronounced $\delta$ while the amplitude
of $\theta$ remains low due to very low number of oscillators
oscillating in synchrony in the C ensemble (see Fig.
\ref{compare_R1}).

On the other hand, for the thalamus (TC+RE) to be synchronized
with the cortex, a relatively higher value of coupling strength
is required. This occurs at $\sim$ $45$ min, at which point the
TC and RE frequencies shift to a higher value to join the C
ensemble and produce $\theta$ activity. This is the reason for
the sudden diminution of the $\delta$ and appearance of
$\theta$ at $\sim$ $45$ min. Of course, the exact time of
occurrence depends on the value of the coupling parameter that
we choose. As a consequence of the shift (inter--ensemble
synchrony), the amount of intra--ensemble synchrony is reduced
in the TC and RE ensembles \cite{Jane:08} and hence the
$\theta$ appears as a low-amplitude activity. We thus quantify
the amplitudes of the characteristic $\delta$ and $\theta$
activities by the mean field parameters $r_{(1,2,3)}$ which
reveals the amount of synchrony in each ensemble and hence the
amplitude of the oscillations. The intra--ensemble and
inter--ensemble synchronisation mechanisms discussed here are
generic to systems of coupled oscillator ensembles
\cite{Winfree:80,Kuramoto:84,Okuda:91,
Strogatz:00,Pikovsky:01,Montbrio:04,Daido:06,Kiss:08,Jane:08}.

\subsection{Phase coding and depth of an{\ae}sthesia}

The $\delta$ and $\theta$ phases play significant roles in
coding information when the brain is not in an aroused state,
characterized by desynchronized EEG. The level with which the
phases are synchronized determines the ability to code
information. However, it is the inter--ensemble synchrony that
plays the crucial role in temporal binding (rather than the
intra--ensemble synchrony). Since synchrony is supposed to
enhance the significance of responses of the neurons
\cite{Singer:99}, it is obvious that synchronized discharges
will have a stronger impact than temporally disorganized ones
\cite{Alonso:96}. Thus the $\theta$ phase is found to play a
crucial role in inhibiting information by being poorly
synchronized in the cortex. More importantly though, the cortex
and the thalamus are not in synchrony with each other (see Fig.
\ref{compare_R1}). Due to this, binding of information cannot
be achieved, which is why consciousness and cognition are
absent during the deep phase of an{\ae}sthesia. The strong
$\delta$ waves keep the cortex and the thalamus out of phase
with each other during deep an{\ae}sthesia. On the other hand,
the thalamus enters into synchrony with the cortex on emergence
from an{\ae}sthesia. This corresponds to the state of awareness
when information can be successfully coded due to the emergence
of inter--ensemble synchrony, characterized by the $\theta$
phase (see Figs. \ref{compare_F1} and \ref{compare_R1}).

\section{Discussion}

The model results indicate that the $\delta$ and the $\theta$
activities observed experimentally are separate in terms of
their generation and frequency. The $\delta$ and $\theta$
activities are found to occur mainly in $0.5-3.5$ Hz and
$3.5-7.5$ Hz bands in agreement with recent observations
\cite{Musizza:07} and earlier reports
\cite{Amzica:98,Amzica:02}. The dramatic diminution of the
$\delta$ amplitude and the simultaneous appearance of $\theta$
activity characterize the transition from deep to light
an{\ae}sthesia.

In the experiment \cite{Musizza:07}, the amplitudes of the
$\delta$ and $\theta$ waves differed by more than a factor of
10. In order to reveal $\theta$ (which otherwise would have
been lost in the noise level of $\delta$) two separate figures
were therefore plotted with different amplitude scales. Our
model results suggest that $\theta$ activity is present during
both deep and light an{\ae}sthesia. However, during deep
an{\ae}sthesia the $\delta$ activity is highly synchronized,
whereas $\theta$ activity is poorly synchronized (see Fig.
\ref{compare_R1}). The effect on their relative amplitudes is
that the $\theta$ activity cannot be seen during the deep
an{\ae}sthesia in the experiment, consistent with the
observations, despite its presence as revealed by the model.

Although varying the coupling parameters to mimic the effect of
decreasing concentration of the an{\ae}sthetic agent helps us
to understand and compare the model results with reality, it
introduces oscillations in the frequencies. As is evident from
Fig. \ref{compare_F1}, the mean frequencies of each of the
ensembles undergo noisy oscillations, while remaining within
the $\delta$/$\theta$ bands. The effect of varying coupling
parameters affects the frequencies directly because the model
equations are for the oscillator phases and the frequencies are
calculated from the phases themselves. Likewise, the
oscillations seen in the order parameters (Fig.
\ref{compare_R1}) are introduced mainly by the change in
inter--ensemble coupling parameters.

Since the model considers only the phases, and not the
amplitudes, we are unable to make a direct comparison with the
experimental results. However, we make use of the theory of
synchronization to quantify the amplitude in terms of the
amount of synchrony in each of the ensembles. That is, the more
the synchrony, the more oscillators are oscillating in phase
with each other, and hence the higher the corresponding
oscillation amplitude. On the other hand, the model does offer
the advantage that we are able to identify those neuronal
groups that are responsible for the generation of $\delta$ and
$\theta$ activity during an{\ae}sthesia.

Although we do not challenge the importance of the microscopic
details underlying general an{\ae}sthesia and the
Hodgkin-Huxley formalisms \cite{Hodgkin:49}, the model results
support the hypothesis that consideration of macroscopic
dynamics with asymmetry is important. We therefore suggest that
models of similar kinds can be used to explain experimental
observations, not only in an{\ae}sthesia, as here, but also in
other cognitive and behavioral states that involve huge numbers
of neurons functioning in groups. Detailed work has been done
in the field of neuronal mass modeling and a wide range of EEG
generative models have been proposed
\cite{Wilson:72,Nunez:74,Silva:76,Freeman:78,Jansen:95,Valdes:99,David:03}.
Our model is capable of generating the wide range of EEG
oscillatory behaviors reported earlier
\cite{Jirsa:03,Wright:01,Wright:03,Rennie:02,Robinson:03}. Our
findings uniquely identify the link between neuronal synchrony
and temporal coding especially in terms of inter--ensemble
synchrony. An additional advantage of the model is that it
yields insight into the synchronization mechanisms underlying
the $\delta$ and $\theta$ waves.

\section{Conclusion}

In summary, we have introduced a model involving asymmetrically
interacting ensembles of C, TC and RE neurons in order to
understand the mechanisms underlying the generation of $\delta$
and $\theta$ waves during an{\ae}sthesia. The model results are
compared with those from experiments \cite{Musizza:07}. The TC
and the RE ensembles were found to be responsible for the
generation of high amplitude $\delta$ waves during deep
an{\ae}sthesia. The C ensemble is engaged with the $\theta$
activity. The transition from deep to light an{\ae}sthesia is
found to be marked by a frequency shift in the TC and RE
ensembles, caused by the increase in the coupling strengths.
Also, the $\theta$ activity is found not to be as strongly
synchronized as the $\delta$ activity. The similarities and
differences between the model results and the experimental
results were discussed. Furthermore, the model illuminates the
phenomenon of temporal coding of information, in particular by
the $\delta$ and $\theta$ frequencies. It reveals the role
played by the neuronal phases in inhibiting information during
deep an{\ae}sthesia and coding the sensory information during
light an{\ae}sthesia. Although our main motivation for
introducing the model was to understand the mechanisms giving
rise to the generation of $\delta$ and $\theta$ waves during
ketamine--xylazine an{\ae}sthesia in rats
\cite{Musizza:07,Stefanovska:00a}, it can also be used to
elucidate the mechanisms involved in the generation of brain
waves for other an{\ae}sthetics and also during slow wave
sleep. The results derived from the model will have
implications for the understanding of fMRI and MEG dynamics.

The study was supported by the EC FP6 NEST-Pathfinder project
BRACCIA and in part by the Slovenian Research Agency.

%
%


\clearpage

\section*{Figure Legends}
\subsubsection*{Figure~\ref{schematic}.}
Left -- The arrangements and connectivity of four types
of cells, namely the TC, RE, C (comprised of the IN and PY
cells). Pre refers to external pre--thalamic afferent sensory
inputs. The upward and downward arrows represent the
thalamocortical and corticothalamic fibres respectively. After
\cite{Destexhe:03}, with permission. Right -- Schematic
representation of the model. There are three ensembles:
cortical neurons(C), thalamocortical relay neurons (TC), and
thalamic reticular neurons (RE). They interact both among
themselves (represented by the circle in each case) and between
each other (represented by the arrows). The upward and the
downward arrows indicate the thalamocortical and
corticothalamic connections. External afferent (pre--thalamic)
inputs to the thalamus are denoted as Pre.

\subsubsection*{Figure~\ref{compare_F1}.}
Left -- The time evolution of the characteristic EEG
$\delta$ and $\theta$ frequencies during Ketamine--xylazine
an{\ae}sthesia, analyzed by wavelet transform. Reproduced from
\cite{Musizza:07}, with permission. Right -- The time evolution
of the characteristic $\delta$ and $\theta$ frequencies
displayed by three ensembles C (blue), TC (red) and RE (green), as
obtained from the model. The (starting) values of the (coupling)
parameters were $A_c=0.8$, $B_c=1.2$, $A_{tc}=0.9$,
$B_{tc}=0.45$, $C_{tc}=0.9$, $A_{re}=0.2$, $B_{re}=0.65$,
$\alpha=0.9$, $D_1=0.1$, $D_2=0.2$, $D_3=0.15$ and
$\gamma=0.4$.

\begin{figure}
\begin{center}
\includegraphics[width=12cm, height=6cm]{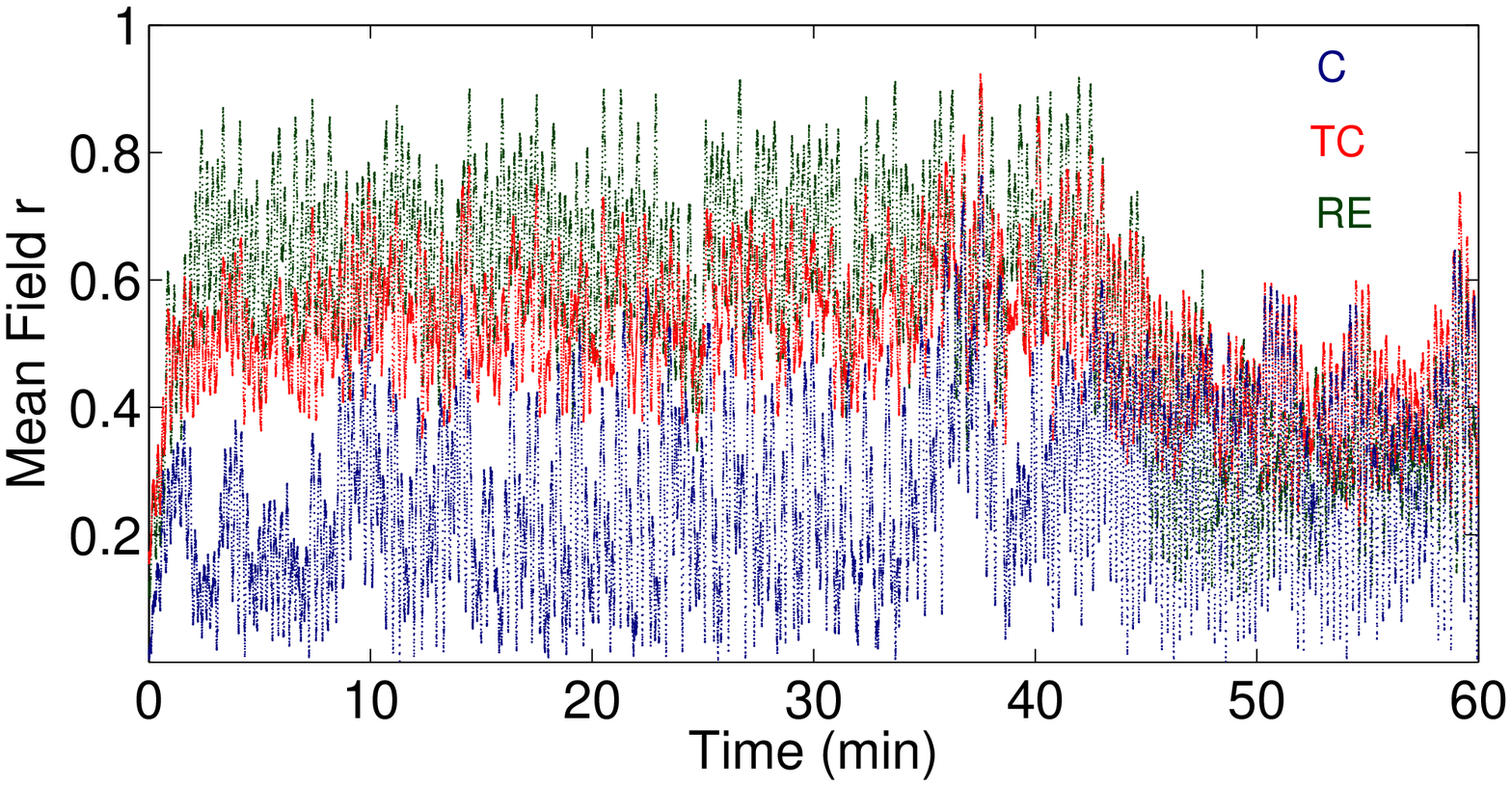}
 \caption{}
\label{compare_R1}
\end{center}
\end{figure}

\subsubsection*{Figure~\ref{compare_R1}.}
\begin{figure}
\includegraphics[angle=0, width=6.5cm]{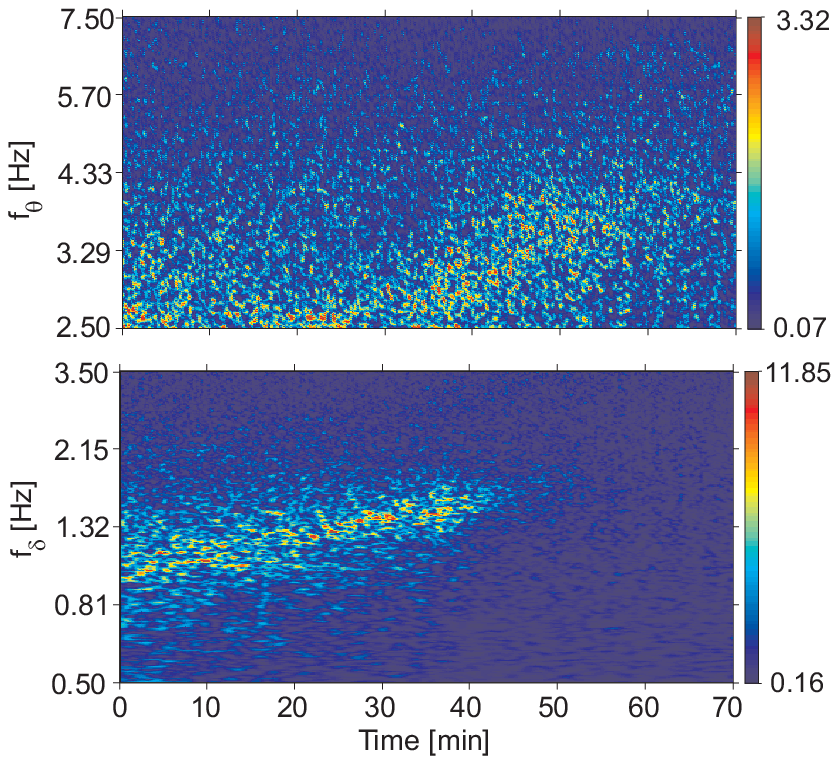}
\includegraphics[angle=0, width=7cm, height=6cm]{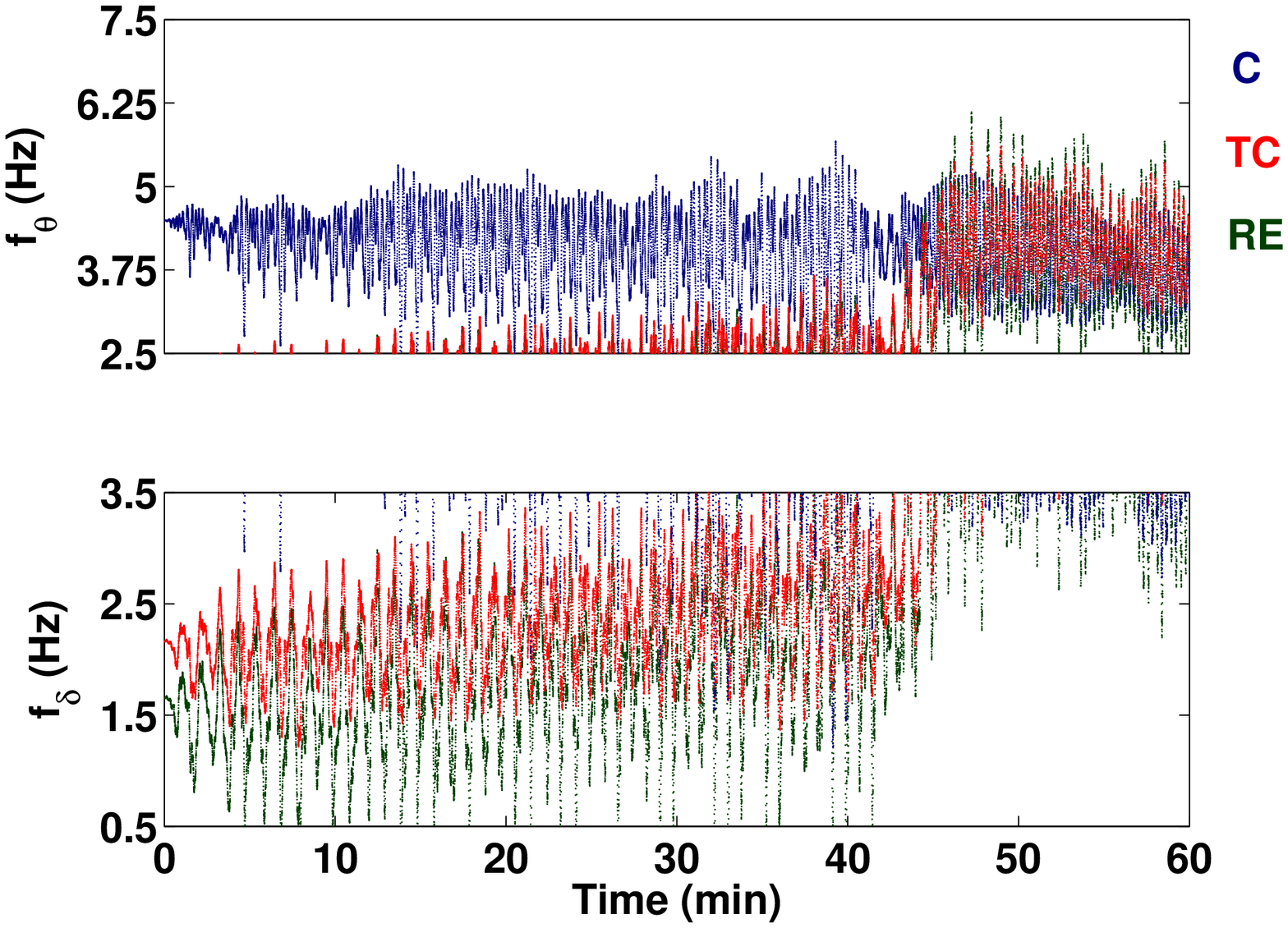}
 \caption{}
\label{compare_F1}
\end{figure}

Mean field parameters $r_{(1,2,3)}$ of the three ensembles, plotted
as functions of time, corresponding to the frequencies plotted in Fig.
\ref{compare_F1} (right) for the same values of parameters.

\clearpage
\clearpage

\end{document}